# Bursting of columnar structures in forced rotating turbulence[*]

Arupjyoti Das[1], Manohar Sharma[2], Avishek Ranjan[1], and Mahendra K Verma[3]

[1]*Department of Mechanical Engineering,*
*IIT Bombay, Mumbai 400076, India*

[2]*LEGI, University of Grenoble Alpes,*
*G-INP, CNRS, Grenoble 38100, France and*

[3]*Department of Physics, IIT Kanpur, Kanpur 208016, India*

(Dated: August 25, 2025)

## Abstract

In this study, we report intermittent bursting of cyclonic columnar structures in direct numerical simulations of forced rotating turbulence in a three-dimensional (3D) periodic box. Columnar structure formation is associated with dominant inverse cascade of energy in addition to the typical forward cascade at large wavenumbers. The forcing is random and isotropic, applied in two distinct wavenumber bands ($k_f = [3, 4]$ and $[6, 7]$). Notably, bursting is observed only for small forcing wavenumber, $k_f = [3, 4]$ and not at $k_f = [6, 7]$, for the same rotation rate and initial conditions. These bursting events are associated with forward cascade of energy in contrast to the inverse cascade of rotating turbulence, as confirmed from the ring spectrum and time-series of modal kinetic energy of modes with $k_z = 2, 3$. Identifying some of the triads involved in the energy transfer from modal analysis, we also calculate the mode-to-mode energy transfer. We find that, before the event, the circular cyclonic vortex becomes elliptical due to cyclone-anticyclone interactions leading to elliptical instability. We argue that energy transfer preceding the bursting events to modes with higher $k_z$ triggers the Crow instability. These instabilities ultimately lead to bursting of the columnar vortex, which subsequently forms again due to the external rotation. Our results portray a tug-of-war between the destabilizing effect of elliptical instability and the stabilizing effect of external rotation.

*This is the accepted manuscript of the article published in Physical Review Fluids (Aug 2025). The final version is available at `https://link.aps.org/doi/10.1103/5kw9-kz43`.*

---

[*] Email address for correspondence:avishekr@iitb.ac.in

## I. INTRODUCTION

Rotating turbulence is a complex phenomenon routinely encountered in various applications, such as in the atmosphere and oceans, liquid interiors of planets and stars, vacuum cleaners, and rotating compressors. The strength of rotation is usually defined by the ratio of inertia to Coriolis forces, called the Rossby number, ($Ro = U/2\Omega' L$). Here $U$, $L$ are the characteristic velocity and length scales, and $\Omega'$ denotes the rotation rate. For $Ro \ll 1$, rotating turbulence is distinguished by the presence of an inverse energy cascade at smaller wavenumbers, leading to the formation of quasi-two-dimensional (columnar) flow structures. These columnar structures are reminiscent of Taylor columns, which arise in rotating fluid systems due to the rotational constraint in the Taylor-Proudman theorem [1, 2]. Typically, these structures are predominantly cyclonic, that is, they rotate in the same sense as the background rotation. These vortices have been observed both in experiments [3–7] and in direct numerical simulations [8–14]. The mechanism of structure formation and the resulting anisotropy is usually attributed to nonlinear triadic interactions of inertial waves which selectively transfer energy across scales. This process is statistically governed by the instability assumption[10, 15, 16] which suggests that, within the resonant triads, energy is preferentially transferred away from the mode that is statistically unstable (identified by the relative sign structure of its triad interaction coefficient) to the other two modes. This directional energy transfer promotes the growth of large scale anisotropic structures and supports the emergence of coherent columnar vortices in rapidly rotating flows. However, it was argued by Staplehurst et al. [5] in their experimental work that quasi-linear inertial wave propagation at a timescale of $\Omega^{-1}$ can explain the columnar structure formation, at least in some of the laboratory experiments that have $0.3 < Ro < 1$ [17] and for non-homogeneous layers of turbulence under rotation [18, 19]. The mechanism of structure formation in experiments is not fully understood. In nearly all investigations the columns are long-lived, as long as the rotation is present. In contrast, our study reveals that, under low-wavenumber forcing, these columnar flow structures exhibit abrupt intermittent bursting, followed by structure formation, for several bursting cycles.

In the context of rotating turbulence in a periodic box, both forced [7, 13, 14, 20–24] and decaying numerical simulations [6, 8, 25, 26] have been reported, with some common but a few dissimilar features. In the decaying simulations, the energy decay rate is found to be lower due to the presence of rotation as compared to classical turbulence, and the coherent structures are long-lived [27]. As the Rossby number decreases from $Ro \sim \mathcal{O}(1)$, the system transitions from a forward cascade (of classical turbulence) to a dual cascade, featuring both inverse and forward energy transfers. Due to the inverse cascade, the horizontal size of the columns ($l_\perp$) increases with time. The same transition occurs for forced rotating turbulence as well but here the forcing wavenumber ($k_f$) is an important control parameter apart from the $Ro$ [10, 23, 24, 26]. Moreover, if the $k_f$ is large, the flow structures in forced simulations are not as coherent as those observed in decaying simulations, due to the dual cascade of the energy injected in smaller scales (for instance, see Sharma et al. [28] where $k_f \in [40, 41]$).

To understand the effect of $k_f$ on the structure formation, we force our simulations at two small wavenumber ranges, $k_f \in [3, 4]$, and $k_f \in [6, 7]$. As expected for strong rotation (small $Ro$), the column formation does occur in both cases due to inverse cascade towards an even smaller $k_f$. However, for the case with $k_f \in [3, 4]$, these columns burst into smaller structures at later times, which then regenerate and form a coherent column, only to then burst once again. As of our knowledge, such intermittent bursting of columnar structures has not been widely reported in the literature on rotating turbulence but has been observed in other flows [29–33]. For instance, for rotating Taylor-Green vortex flow, Alexaxis [29] observed periodic energy fluctuations, particularly at low-wavenumber forcing ($k_z = 2, k_\perp = 2\sqrt{2}$). He found that the amplitude of certain unstable modes grows, after which the system transitioned to a highly non-linear state, causing its energy to transfer to all modes and cascade to the dissipation scales. However, the energy in the quasi-two-dimensional modes predominantly remained in large scales, forming a condensate at $k_z = 0$, $k_\perp = 1$, suppressing the instability.

Intermittent bursting behavior has been reported in wall-bounded turbulence [30], in astrophysical flows influenced by precession [31, 32], in tidally deformed bodies undergoing elliptical instability [34] and in laboratory experiments forced with imposed elliptical strain [35]. In these studies, intermittent bursts are caused by the elliptical instability, which emerges from the resonance between inertial waves and elliptical deformation. For instance, Eloy et al. [36] observed similar bursting patterns in experiments conducted within closed rotating cylinders, concluding that bursting occurs due to the resonance of the Kelvin modes with the deformation field. Moreover, when counter-rotating vortex pairs interact, they can trigger instability involving both shortwave and longwave interactions [37–39].

We report here intermittent bursting (and subsequent re-formation) of columnar flow structures in simulations of forced rotating turbulence. We also investigate the underlying mechanism, identifying the crucial system parameters ($k_f$, $\nu$, $\Omega$) that lead to the occurrence of bursting events. To this end, we also study in detail the behavior of Fourier modes during these bursting events.

The paper is structured as follows. First, in section II, we discuss the numerical methodology including governing equations in Fourier space along with the parameters. Thereafter, in section III, we report our findings in terms of 3D visualizations, time-series of modes, modal energy transfer and ring spectrum. We discuss our findings in the context of literature on elliptical instability in III C before we finally conclude in section IV.

## II. NUMERICAL METHODOLOGY AND SIMULATION DETAILS

### A. Governing Equations

In our numerical simulations, we use the dimensionless form of the governing equations in the rotating frame with the following dimensionless variables: $x = x'/L$, $\mathbf{u} = \mathbf{u}'/U$, $t = t'/(L/U)$, $\nu = \nu'/(UL)$, $\nabla = L\nabla'$, $p = p'/(\rho U)^2$, $\mathbf{\Omega} = \mathbf{\Omega}'L/U$ . Here, rotation is present along the z-axis, $L$ is the system size (we assume a cubical domain in our case) and $U$ is a characteristic velocity scale, and $'$ denotes the dimensional quantities. Along with the continuity equation,

$$\nabla \cdot \mathbf{u} = 0, \tag{1}$$

the (dimensionless) Navier-Stokes (NS) equation in rotating reference frame, therefore, is:

$$\frac{\partial \mathbf{u}}{\partial t} + (\mathbf{u} \cdot \nabla)\mathbf{u} = -\nabla p - 2\mathbf{\Omega} \times \mathbf{u} + \nu \nabla^2 \mathbf{u} + \mathbf{f}. \tag{2}$$

Note that $\nu$ and $2\Omega$ represent the inverse of Reynolds and Rossby numbers based on characteristic scales, respectively.

In Fourier (spectral) space, the velocity field is expanded as $\mathbf{u}(\mathbf{x}, t) = \sum_{\mathbf{k}} \hat{\mathbf{u}}(\mathbf{k}, t)e^{i\mathbf{k} \cdot \mathbf{x}}$, where $\hat{\mathbf{u}}(\mathbf{k}, t)$ is the Fourier coefficient corresponding to the wavenumber $\mathbf{k}$. The governing equations for incompressible flow in a rotating frame then become:

$$\mathbf{k} \cdot \hat{\mathbf{u}}(\mathbf{k}, t) = 0, \tag{3}$$

$$\frac{d\hat{\mathbf{u}}(\mathbf{k}, t)}{dt} = -i\mathbf{k}\hat{p}(\mathbf{k}, t) - \hat{\mathbf{N}}_u(\mathbf{k}, t) - 2\mathbf{\Omega} \times \hat{\mathbf{u}}(\mathbf{k}, t) - \nu k^2 \hat{\mathbf{u}}(\mathbf{k}, t) + \hat{\mathbf{f}}(\mathbf{k}, t). \tag{4}$$

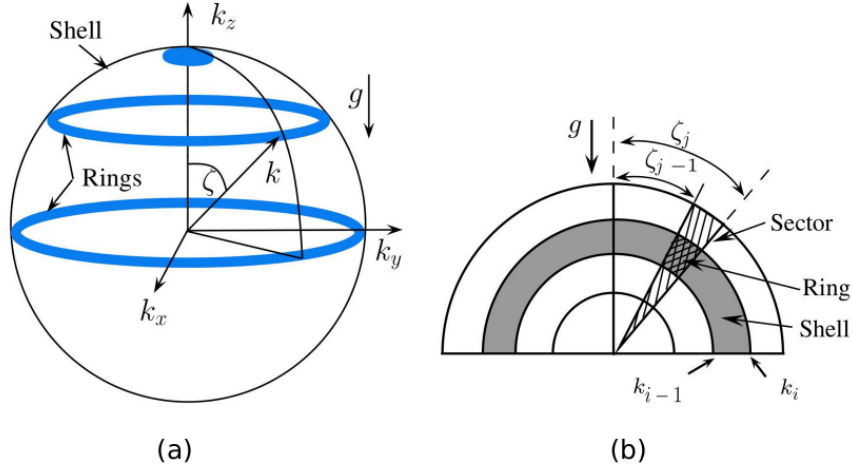

FIG. 1. (a) Schematic diagram exhibiting spherical rings in wavenumber space (b) Vertical cross-section showing ring sector and shell, reproduced with permission from Verma [40].

Here $\hat{N}_u(\mathbf{k}, t) = i \sum_p [\mathbf{k} \cdot \hat{\mathbf{u}}(\mathbf{q})] \cdot \hat{\mathbf{u}}(\mathbf{p})$ is the non-linear advection term in the NS equation, where $\mathbf{q} = \mathbf{k} - \mathbf{p}$. More details about the spectral method and relevant tools can be found in Verma [40] but we include them here for completeness.

The formation of coherent vortex columns in rotating turbulence enhances anisotropy. In order to quantify the anisotropy in the distribution of energy across Fourier modes, we can use the ring spectrum, where the Fourier space is decomposed into shell and sectors.

A shell with index $i$ in Fourier space is defined as:

$$\text{shell}(i) := \{\mathbf{k} : \mathbf{k}_{i-1} \leq |\mathbf{k}| < \mathbf{k}_i\} \tag{5}$$

where $i \in \mathbb{N}$, and $\mathbf{k}_{i-1}$ and $\mathbf{k}_i$ are the radii of the $(i-1)^{\text{th}}$ and $i^{\text{th}}$ concentric spheres in Fourier space. The shells essentially partition the Fourier space into non-overlapping concentric annular regions such that their union encompasses the entire Fourier space under consideration.

A sector with sector index $j$ in Fourier space is denoted by sector($j$) and can be defined [28]:

$$\text{sector}(j) := \{\mathbf{k} : \zeta_{j-1} \leq \arccos\left(\frac{k_\parallel}{|\mathbf{k}|}\right) < \zeta_j\} \tag{6}$$

where $\mathbf{k}_\parallel$ is the component of the wave vector along the direction of the rotation axis, i.e $z$-axis and $\zeta_j \in [0, \pi]$ is the angle between the $z$-axis and the corresponding wave vector. A sector is a volume entrapped between the two polar angles $\zeta_{j-1}$ and $\zeta_j$.

Any ring is the intersection between a shell and a sector. In general, a ring of index $(i, j)$ is defined as [40]:

$$\mathrm{ring}(i, j) := \{\mathbf{k} : \mathbf{k} \in \mathrm{shell}(i) \cap \mathrm{sector}(j)\}, \tag{7}$$

where $i, j \in \mathbb{N}$. The rings lie inside the angular range from $\zeta = 0$ to $\zeta = \frac{\pi}{2}$ due to $\zeta = 0 - \pi$ symmetry. The shell, sector and the ring are illustrated in Fig. 1. Following Reddy and Verma [41], the kinetic energy spectrum of ring $(\mathbf{k}, \beta)$, with shell$(\mathbf{k})$ and sector$(\beta)$ in Fourier space may be defined as:

$$E_u(\mathbf{k}, \beta) = \frac{1}{\mathrm{Norm}(\mathbf{k}, \beta)} \sum_{\substack{\mathbf{k}-1 < \mathbf{k}' \leq \mathbf{k} \\ \beta \in (\zeta_{\beta-1}, \zeta_\beta)}} \frac{1}{2} |\hat{\mathbf{u}}(\mathbf{k}')|^2. \tag{8}$$

The normalizing constant is used to compensate for the uneven distribution of modes in the sector, where a sector near the equator contains relatively larger number of modes than at the poles, is given by:

$$\mathrm{Norm}(\mathbf{k}, \beta) = |\cos(\zeta_{\beta-1}) - \cos(\zeta_\beta)|. \tag{9}$$

The ring spectra for the perpendicular and parallel components of the energy are defined as:

$$E_\perp(\mathbf{k}, \beta) = \frac{1}{\mathrm{Norm}(\mathbf{k}, \beta)} \sum_{\substack{\mathbf{k}-1 < \mathbf{k}' \leq \mathbf{k}, \\ \angle(\mathbf{k}') \in (\zeta_{\beta-1}, \zeta_\beta)}} \frac{1}{2} |\hat{\mathbf{u}}_\perp(\mathbf{k}')|^2, \tag{10}$$

$$E_\parallel(\mathbf{k}, \beta) = \frac{1}{\mathrm{Norm}(\mathbf{k}, \beta)} \sum_{\substack{\mathbf{k}-1 < \mathbf{k}' \leq \mathbf{k}, \\ \angle(\mathbf{k}') \in (\zeta_{\beta-1}, \zeta_\beta)}} \frac{1}{2} |\hat{\mathbf{u}}_\parallel(\mathbf{k}')|^2. \tag{11}$$

To gain a deeper understanding of the formation of large coherent structures and their subsequent fragmentation into smaller structures, a detailed examination of energy transfer between the modes is insightful. Dar *et al.* [42] derived the mode-to-mode energy transfer denoted as $S(\mathbf{k}|\mathbf{p}|\mathbf{q})$ for a triad $(\mathbf{k}, \mathbf{p}, \mathbf{q})$ where $\mathbf{k} = \mathbf{p} + \mathbf{q}$.

$$S(\mathbf{k}|\mathbf{p}|\mathbf{q}) = \Im\left[(\mathbf{k} \cdot \hat{\mathbf{u}}(\mathbf{q}, t))(\hat{\mathbf{u}}(\mathbf{p}, t) \cdot \hat{\mathbf{u}}^*(\mathbf{k}, t))\right]. \tag{12}$$

In Fourier space, the mode-to-mode energy transfer represents the rate at which energy is transferred from the velocity mode $\hat{\mathbf{u}}(\mathbf{p})$ to the velocity mode $\hat{\mathbf{u}}(\mathbf{k})$ through the intermediary velocity mode $\hat{\mathbf{u}}(\mathbf{q})$. This technique allows us to discern the role of each mode in the energy transfer process, providing valuable insights into the anisotropic behavior of the system [42, 43].

## B. Numerical Methodology

We solve the governing equations (Eqs. 3 and 4) using the pseudo-spectral DNS code TARANG [44] in a 3D periodic cubic domain of size $(2\pi)^3$. The code utilizes a fourth-order Runge-Kutta scheme for time integration, adhering to the Courant–Friedrichs–Lewy (CFL) condition to determine an optimal time step ($\Delta t$), with the 2/3 rule for dealiasing. In most of our simulations, pre-dealiased grid resolution is $256^3$ which helps us visualize the results in 3D. To achieve a steady-state flow, a random forcing technique on the velocity field [45–47] is employed, a method proposed by Carati et al. [45]. This forcing approach is designed to provide a constant energy supply to the flow while maintaining zero kinetic helicity. The energy supply rate, denoted as $\epsilon_{\text{in}}$, is distributed among $n_f$ Fourier modes within the defined forcing shell.

The amplitude of the forcing, $\hat{\mathbf{f}}(\mathbf{k})$, is expressed as:

$$\hat{\mathbf{f}}(\mathbf{k}) = \frac{\epsilon_{\text{in}} \hat{\mathbf{u}}(\mathbf{k})}{n_f \left( \hat{\mathbf{u}}(\mathbf{k}) \cdot \hat{\mathbf{u}}^*(\mathbf{k}) \right)}, \qquad (13)$$

where the operator $(\cdot)$ denotes the scalar product. It is important to note that this forcing is inherently random, as the velocity $\hat{\mathbf{u}}(\mathbf{k})$ exhibits random phase characteristics.

To establish the initial conditions for simulating rotating turbulent flows, we first conducted a numerical simulation of a non-rotating turbulent flow with a $256^3$ grid resolution with $\nu = 0.005$ (it is useful to recall here that $\nu$, and $\Omega$ are dimensionless). Within this setup, we apply forcing in the $k \in [3, 4]$. This forcing ensures a constant kinetic energy supply rate of $\epsilon_{\text{in}} = 0.4$ within the flow. To initiate the simulation of rapidly rotating turbulence, we use the steady-state data obtained from fully developed hydrodynamic turbulence at $t = 45$ where $\epsilon_{\text{in}} = \epsilon$.

In our simulations, the dimensionless control parameters are the (forcing scale) Reynolds ($Re_f$) and Rossby number ($Ro_f$) [23, 24] defined as,

$$Ro_f = \frac{U_f}{\Omega L_f} \quad \text{and} \quad Re_f = \frac{U_f L_f}{\nu}. \qquad (14)$$

The values of $Ro_f$ and $Re_f$ are mentioned in Table I. Here, the forcing length scale $L_f = \frac{2\pi}{k_f}$ represents the characteristic scale at which energy is injected, determined by the forcing wavenumber $k_f$ (mid-value of the forcing range) and $U_f = (\epsilon_{\text{in}} L_f)^{1/3}$ reflects the typical velocity associated with turbulence at this scale, all of which are kept fixed in our simulations.

TABLE I. Parameter values for different simulations

| Case | $k_f$ | Resolution | $\nu$ | $\Omega$ | $Ro_f$ | $Re_f$ | $Ro_L$ | $Re_L$ |
|---|---|---|---|---|---|---|---|---|
| 1 | [3,4] | $(256)^3$ | 0.01 | 16 | 0.031 | 160.7 | 0.008 | 485.99 |
| 2 | [3,4] | $(256)^3$ | 0.005 | 16 | 0.031 | 321.51 | 0.008 | 971.97 |
| 3 | [3,4] | $(256)^3$ | 0.005 | 8 | 0.062 | 321.51 | 0.015 | 971.97 |
| 4 | [3,4] | $(256)^3$ | 0.005 | 24 | 0.021 | 321.51 | 0.005 | 971.97 |
| 5 | [6,7] | $(256)^3$ | 0.005 | 16 | 0.047 | 70.42 | 0.005 | 328.8 |
| 6 | [6,7] | $(256)^3$ | 0.01 | 16 | 0.047 | 140.84 | 0.005 | 656.06 |
| 7 | [6,7] | $(512)^3$ | 0.005 | 16 | 0.047 | 140.84 | 0.008 | 972 |

For comparison, in Table I, we also show the (classical) Rossby number ($Ro_L$) and Reynolds number ($Re_L$) at the start of the rotating runs i.e. at $t = 0$, these are defined as:

$$Ro_L = \frac{U_{rms}}{\Omega L} \quad \text{and} \quad Re_L = \frac{U_{rms} L}{\nu}. \tag{15}$$

Here, $U_{rms}$ represents the r.m.s. velocity expressed as $U_{rms}(t) = \left(\frac{2}{3}E(t)\right)^{1/2}$, where $E(t)$ is the volume-averaged energy of the system and $L$ is the box length. We force our simulations in the two wavenumber ranges, $k_f = [3,4]$ and $k_f = [6,7]$, to assess the effect of forcing scale. To check grid-independence we have performed one simulation at a higher resolution of $512^3$ and the time-series of kinetic energy and dissipation rates (not reported here) are observed to exactly overlap. We have also validated the code using two test cases, first with the Taylor-Green vortex [48] as well as using a Gaussian vortex under the effect of rotation [11].

## III. RESULTS AND DISCUSSION

The time-series of the volume-averaged kinetic energy of the system ($E(t)$), and dissipation rate ($\epsilon$), are shown in Fig. 2. An increase in $E(t)$ is observed in Fig. 2(a), when background rotation is introduced. This behavior is expected as it is well-established that rotation causes a reduction in the rate of energy dissipation, apart from the inverse cascade leading to the columnar structure formation [27]. Additionally, for a constant rotation rate, Fig. 2(a) demonstrates a decrease in $E(t)$ with increasing viscosity ($\nu$), as indicated by the dashed line. This can be attributed to the fact that, for large-scale columnar vortices resembling quasi-two-dimensional flows, the dissipation rate per unit mass scales as $\epsilon_{2D} \sim \frac{\nu U^2}{L^2}$.

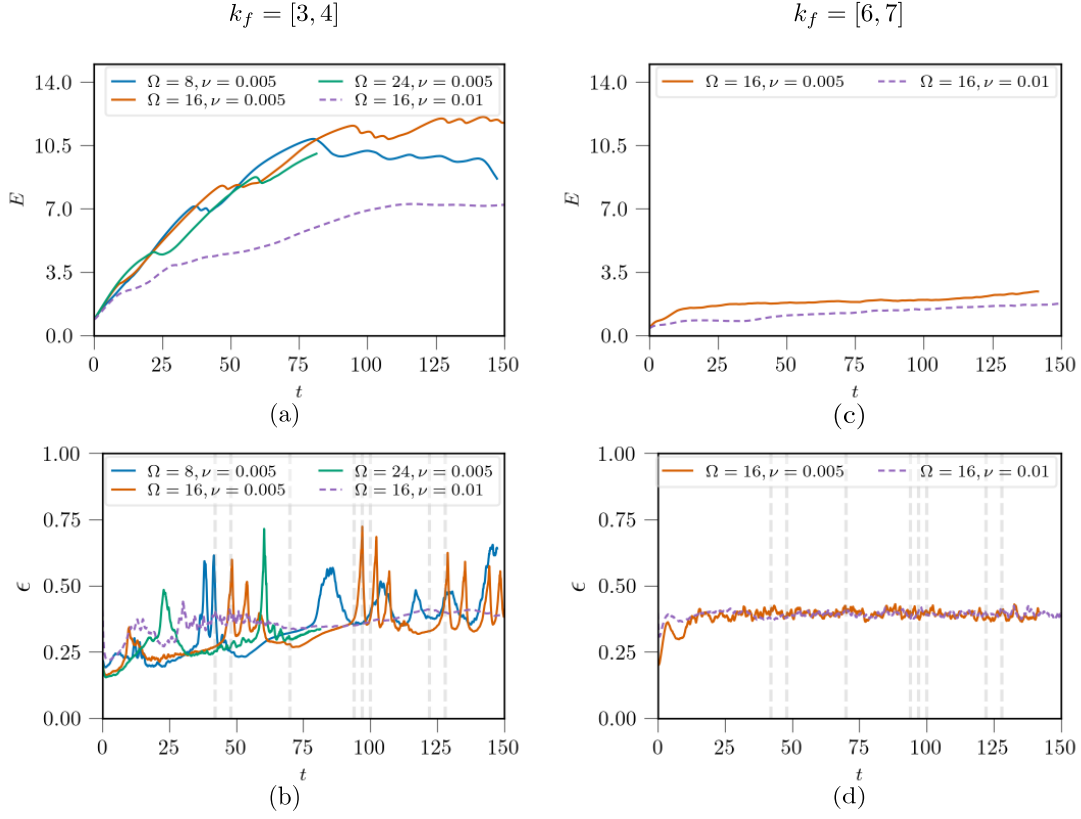

FIG. 2. Time evolution of (a) $E$ and (b) $\epsilon$ at three distinct rotation rates ($\Omega = 8, 16, 24$) for forcing wavenumber $k_f = [3, 4]$ (left column). Time evolution of (c) $E$ and (d) $\epsilon$ for a fluid system with rotation rate ($\Omega = 16$) for forcing wavenumber $k_f = [6, 7]$ (right column). Solid and dashed lines are for two viscosities ($\nu = 0.005$ and $\nu = 0.01$) respectively.

We find an intriguing observation in Fig. 2(b): intermittent peaks (or spikes) are seen in the dissipation rate ($\epsilon$) for the case with $\nu = 0.005$ that are absent in the $\nu = 0.01$ case. As will be shown later, these spikes are due to the bursting of columnar structures. During these bursting events, a sudden unexpected increase in the dissipation rate occurs, leading to fluctuations in the overall energy dissipation. The effect of increase in rotation rate on this phenomenon is notable, in that it stabilizes the system and delays the occurrence of these intermittent bursts.

In Fig. 2(c) and Fig. 2(d), we plot the $E(t)$ and $\epsilon(t)$ for a higher forcing wavenumber,

$k_f = [6, 7]$. Notably, although the $E(t)$ in the system still increases, the rate of increase and the maximum energy are lower as compared to the lower $k_f$ case. Unlike the lower $k_f$ case, there are no fluctuations in the $E(t)$ and the dissipation rates. These spikes observed at lower $k_f$ are of particular interest, and a detailed investigation is conducted to explore the underlying reasons for their presence.

We investigate the structure and statistical properties of the system for time frames reflected by the grey dashed vertical lines in Fig. 2(b) and Fig. 2(d). We choose to focus on the results from the cases 2 and 5 in the following sections, both of which are at small $Ro_f$, but at different $k_f$.

### A. Bursting of Coherent Structures

The evolution of the z-component of vorticity ($\omega_z$) with time for case 2 is presented in Fig. 3. At initial times, the maximum of cyclonic and anticyclonic vorticity is nearly equal. Subsequently, we observe that the small vortices merge to form a single large coherent structure (Fig. 3(b)) [6, 49]. At $t = 48$ (Fig. 3(c)), we observe an unexpected behavior where the columnar structure bursts into smaller structures. The bursting is characterized by sudden spikes in the energy dissipation rate shown earlier in Fig. 2(c). After the bursting, the columnar flow structure forms once again. This cycle of intermittent bursting followed by column formation repeats again at $t = 97$ and $t = 128$. We also observe that the maximum $\omega_z$ acquires its peak value at the time of bursting and then decreases. Additionally, prior to the bursting, the structures exhibit a wavy configuration, as seen in Fig. 3(e) and Fig. 3(h) at $t = 42$ and $t = 122$ respectively. We observed similar behavior for other rotation rates as well where spikes in $\epsilon$ are observed.

Interestingly, this disruptive behavior is absent in cases 1 and 6 (see Table I) with $\nu = 0.01$ for both $k_f$ values, as evidenced by the lack of spikes in $\epsilon$ in Fig. 2(b) and Fig. 2(d). Moreover, for the case with $k_f = [6, 7]$ and $\nu = 0.005$, the columns are no longer coherent and they appear to be fragmented along the rotation axis.

In Fig. 4, we exhibit the vertical (xz) cross-section of vorticity ($\omega_z$) through the cyclone and anticyclone pair. We observe that over time the contour of the anticyclonic vortex gradually approaches the cyclonic vortex, leading to their eventual interaction. Notably, in Fig. 4(c), small structures begin to separate from the initially coherent column, leading to the formation of wavy columns observed in Fig. 3(e) and Fig. 3(h). During the bursting event, the peak of the anticyclone moves within the bursting cyclone. This is evident more clearly in a movie attached as a Supplementary material.

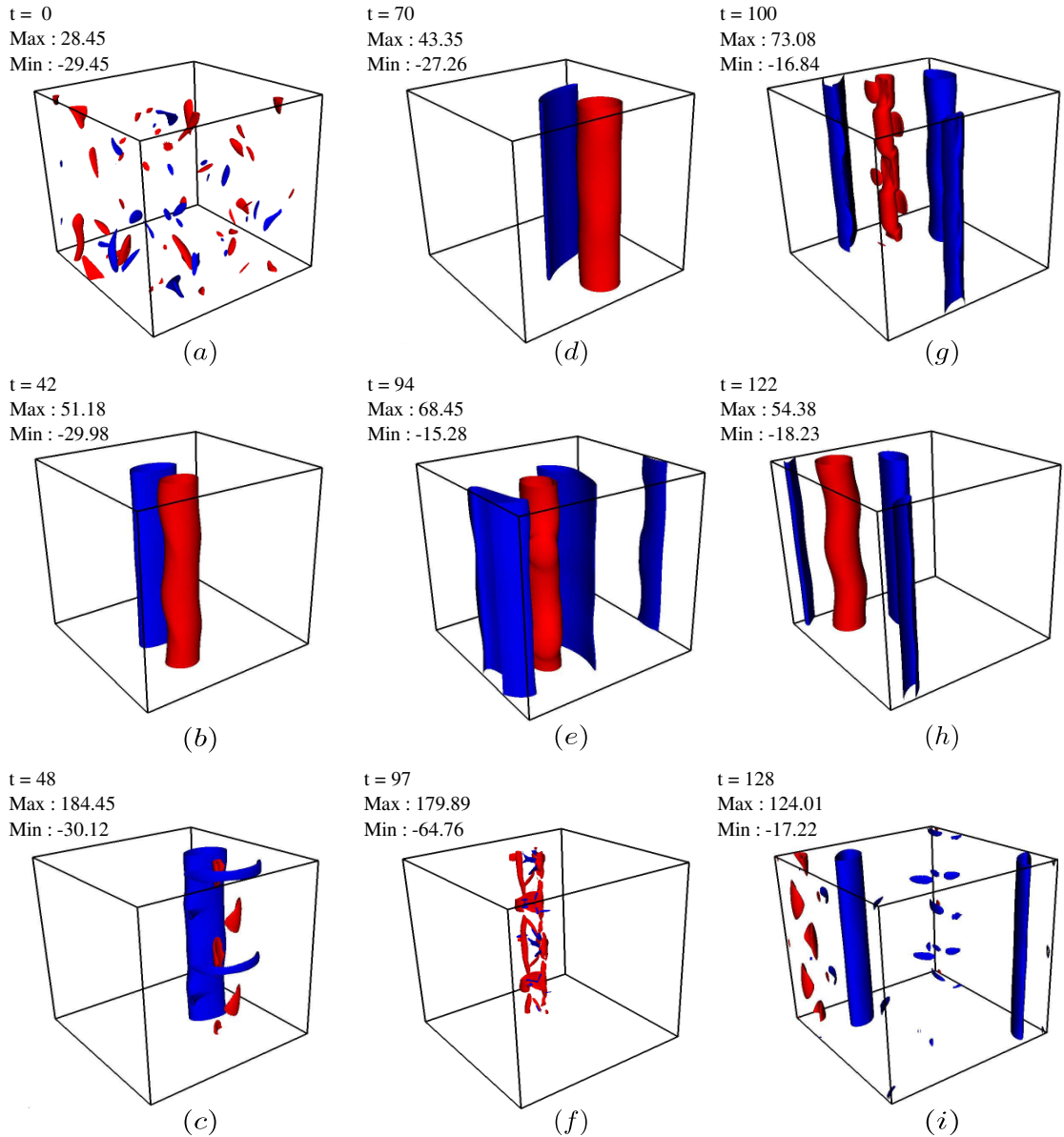

FIG. 3. Vorticity contour evolution for case 2 at times - (a) $t = 0$, (b) $t = 42$, (c) $t = 48$, (d) $t = 70$, (e) $t = 94$, (f) $t = 97$, (g) $t = 100$, (h) $t = 122$, (i) $t = 128$. Red and Blue color represent cyclones and anticyclones, respectively. (Contours at 50% of Maxima and Minima).

According to Laporte and Corjon [50]), the interaction between a pair of counter-rotating Lamb-Oseen vortices, in the absence of global rotation, can lead to bursting phenomena due to the simultaneous emergence of Crow and elliptic instabilities. The strain induced by

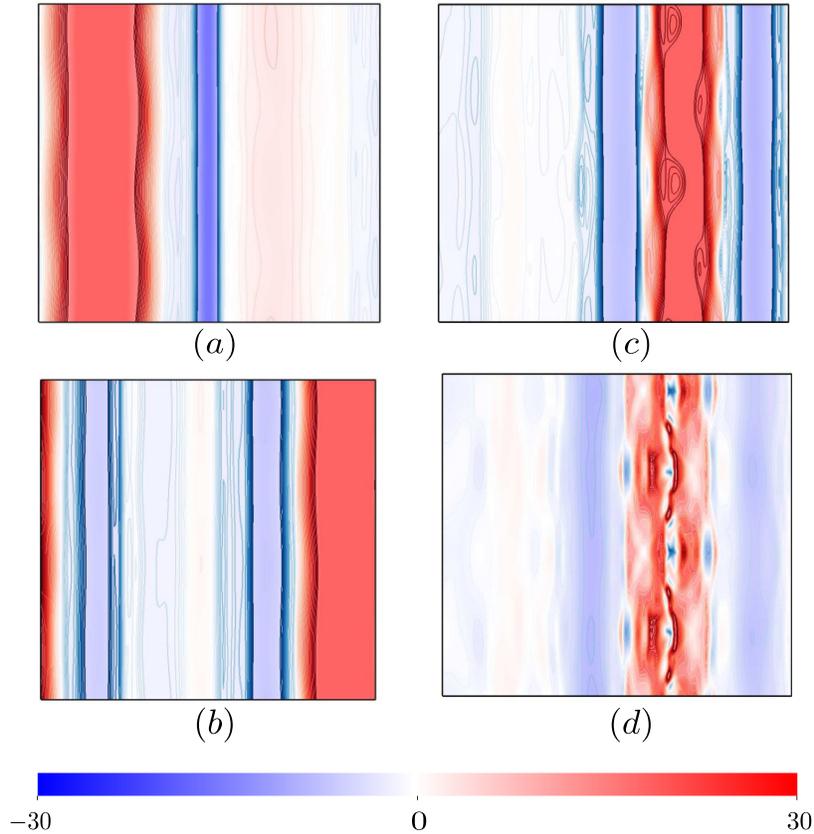

FIG. 4. Vertical Plane of $\omega_z$ through the cyclone and anti-cyclone pair for case 2 at time (a) $t = 70$, (b) $t = 90$, (c) $t = 94$, (d) $t = 97$

.

the neighboring vortex accelerates the development of elliptical streamlines, particularly in regions where the Crow instability reduces the vortex spacing. The elliptic instability breaks down the vortex into smaller-scale flow. Since our simulation also has a cyclone-anticyclone vortex pair, albeit in a rotating frame of reference, it is likely that the same bursting mechanism is prevalent in our simulation. We will elaborate further on this in section III C.

### B. Ring spectrum, modal analysis and energy transfer

An analysis of modes and energy transfer in Fourier space sheds light on the complex bursting phenomenon. Recall that the bursting is observed only for $k_f = [3, 4]$ (case 2) and not for $k_f = [6, 7]$ (case 5). In Fig. 5 we illustrate the energy spectrum for both cases at the

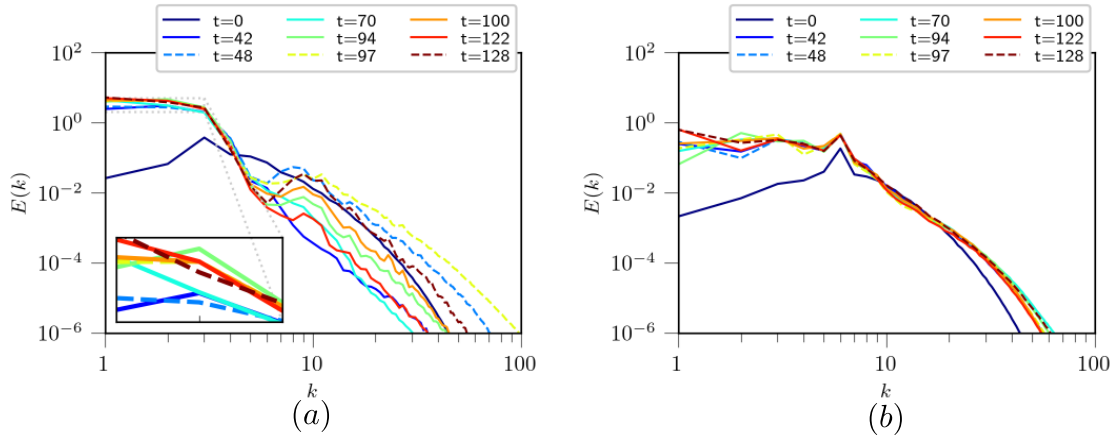

FIG. 5. The energy spectrum at different timeframes for different forcing wavenumber for (a) case 2 ($k_f = [3, 4]$), and (b) case 5 ($k_f = [6, 7]$). Inset presents the zoomed energy spectra in the small wavenumbers

time frames indicated by the grey dashed lines in Fig. 2.

At large k, we observe a notable increase in $E(k)$ for the case 2 (Fig. 5(a)) during the bursting events at $t = 48, 97, 128$, see Figs. 2 and 3. This accumulation of energy at large $k$ vanishes after the bursting period. Interestingly another important observation is that the spectra at $t = 94$ shows a forward cascade of energy with greater energy at higher wavenumbers compared to $t = 70$, when steady coherent columns were present. This is accompanied by the shift of maxima from $k = 1$ to $k = 2$ (inset of Fig. 5(a)). This shift towards $k = 2$ correlates with the formation of the wavy vorticity contours observed at that time. Similar observation can be made for the spectra at $t = 42$. In contrast, for the case 5, at a larger $k_f$, we do not see any notable variation in $E(k)$ after the initial change due to inverse cascade (Fig. 5(b)), a feature typical of rotating turbulence.

TABLE II. Energy distribution in the perpendicular and parallel directions at $t = 90, 97, 100$. Note the values of $E_\parallel$ before and after the bursting event - they are not equal.

| $t$ | $E_{total}$ | $E_\perp$ | $E_\parallel$ |
| --- | --- | --- | --- |
| 90 | 11.41 | 11.40 | 0.01 |
| 97 | 11.30 | 11.20 | 0.10 |
| 100 | 11.26 | 11.23 | 0.03 |

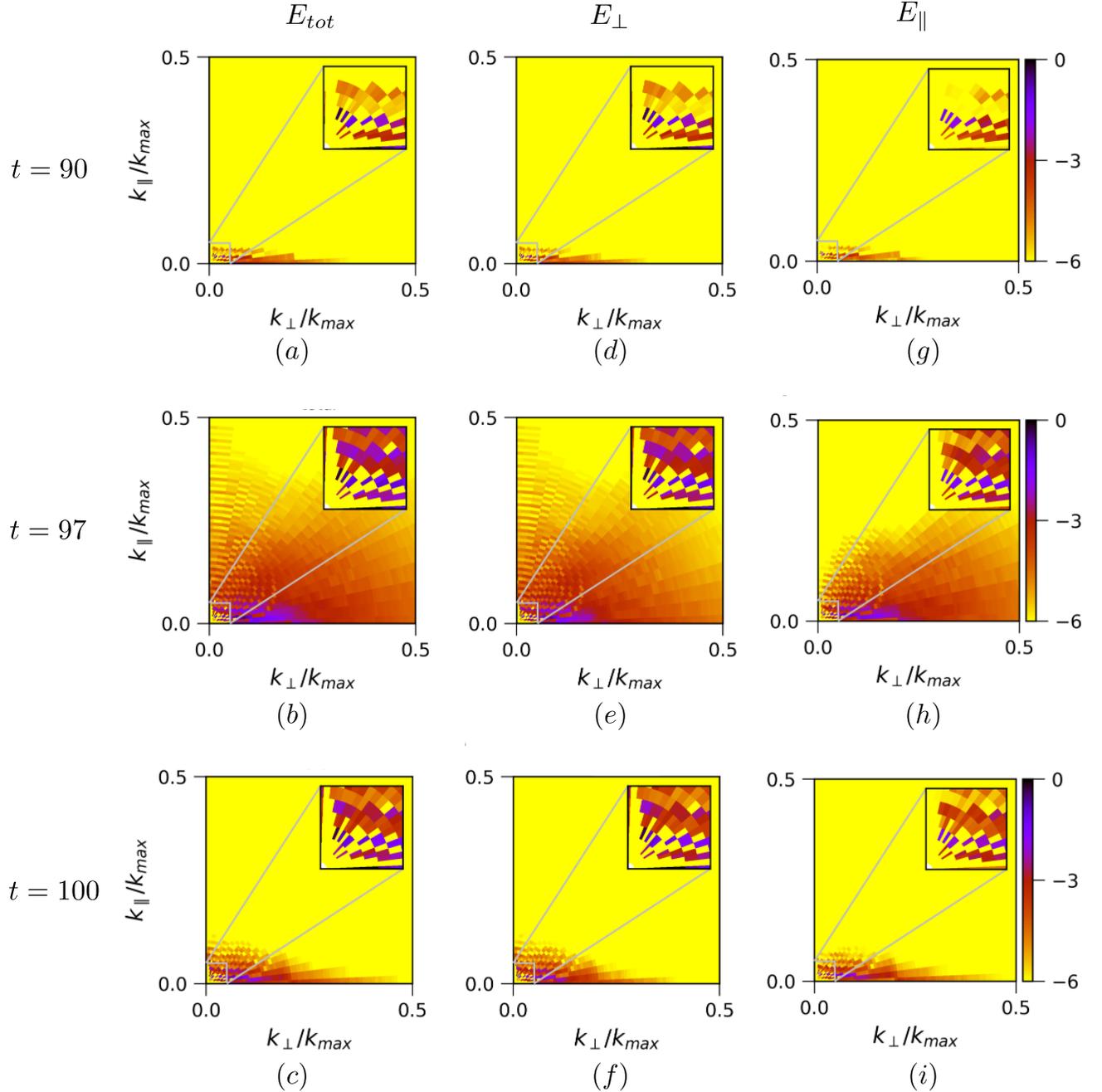

FIG. 6. Anisotropic energy spectra of $E_{total}$ (a-c), $E_\perp$ (d-f), and $E_\parallel$ (g-i) at three different time frames $t = 90, 97$ and $100$ for case 2. Insets are included to present the zoomed energy content in the small wavenumber range.

In Fig. 6, the ring spectrum plots at $t = 90, 97$ and $100$ depicting a particular bursting event are presented. The values of $E_\parallel$, $E_\perp$ and total energy are included in the Table II. At $t = 90$, before the bursting occurs, we observe that most of the total energy is concentrated in wavenumbers perpendicular to the axis of rotation with small $k_\parallel$ [28]. The occurrence of bursting event at $t = 97$ leads to a cascade of energy across a larger range of $k_\parallel$ and $k_\perp$, accompanied by a discernible increase in the energy content of higher $k_\parallel$ and $k_\perp$ modes. Notably, the total $E_\parallel$ during bursting is nearly 100 times the energy at $t = 90$. This change is evident in the insets shown from Fig. 6(a) and Fig. 6(b). Thus, the bursting event tries to enhance the 3D features within the flow. After the bursting subsides and reformation of structure starts taking place, $E_\parallel$ decreases, as observed from $t = 97$ (Fig. 6(h)) towards the state at $t = 100$ (Fig. 6(i)). However, despite the total energy returning to lower $k_\parallel$ modes, a complete restoration to its initial value is not achieved, see Fig. 6(c). This was also observed by Alexakis et al. [29] for rotating Taylor Green flow.

From the undulating 3D cyclonic columns depicted in Fig 3(b), 3(e), and 3(h), as well as the fully burst columns in Fig. 3(c), 3(f), and 3(i), it is evident that modes with $k_z = 2$ could be dominant apart from $k_z = 0$. Additionally, the cross-section in Fig. 4 and the energy spectrum in Fig. 5(a) also support this observation. However, from Fig. 5 and Fig. 6 we find that higher $k_z$ modes also exhibit increased energy during bursting. But, the energy in these modes is considerably smaller compared to the modes with small $k_z$.

In Fig. 7, we have shown the temporal evolution of modal energy of select modes specified by their Cartesian wave vector components, such as $[0, 1, 2], [-1, -1, 2]$ and $[1, -1, 3]$. A discernible spike in energy of these modes is evident during bursting. For mode $[1, -1, 3]$, we observe negligible energy at $t = 90$, when the structure is intact, but from $t = 93$ onwards, it begins to accrue energy, reaching a maximum around $t = 98$, only to dissipate subsequently. For modes $[-1, -1, 2]$ and $[0, 1, 2]$, we again observe that the modal energy increases from $t = 93$ and the peak in energy occurs around the bursting time, approximately at $t = 97$. Based on these observations, it is plausible to assume that energy is initially transferred to $k_z = 2$ modes, inducing a waviness in the structures. Subsequently, this energy is further transferred to higher $k_z$ modes, associated with smaller flow structures, much in contrast to the coherent column. Similar observations exist for different bursting events corresponding to varying rotation rates ($\Omega$) (not shown here). Additionally, while mode $[0, 1, 2]$ is consistently involved in numerous bursting events, the same cannot be asserted for the mode $[1, -1, 3]$.

In simulations of decaying rotating turbulence, Sharma et al. [49] observed that a significant portion of energy is concentrated in the Fourier modes $[\pm 1, 0, 0]$ and $[0, \pm 1, 0]$, which correspond to the largest columnar structures. In our forced simulations, this observation holds true as well, particularly when stable coherent structures are present. Therefore, we hypothesize that a sequential transfer of energy occurs within the modes, from modes with $k_z = 0$ to modes with higher $k_z$, initially inducing waviness and ultimately leading to the bursting into smaller structures.

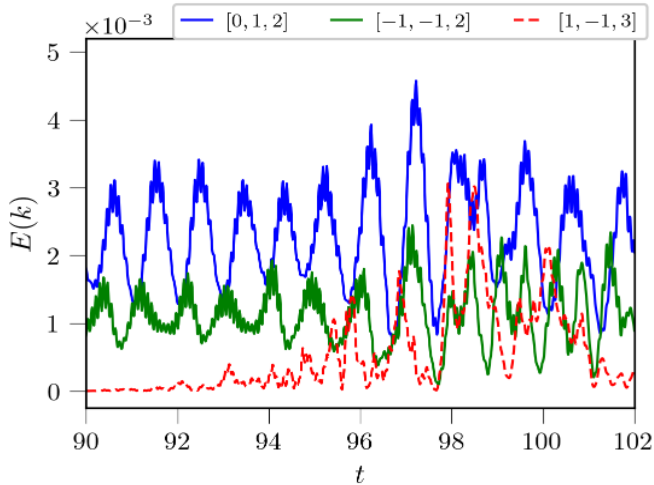

FIG. 7. Temporal evolution of modes for the case 2 during a single bursting event. Dashed line represents the data being multiplied by a factor of 500 for clarity.

In order to substantiate our hypothesis of sequential energy transfer, we now calculate the $S(\mathbf{k}|\mathbf{p}|\mathbf{q})$ corresponding to energy transfer within a triad of modes using Eq. 12 and plot its time evolution in Fig. 8. This offers deeper insights into the intricate dynamics leading up to and during the bursting events.

We focus on the modes that satisfy the criteria for interacting triads i.e. $\mathbf{k} = \mathbf{p} + \mathbf{q}$ where the transfer of energy occurs from the mode $\mathbf{p}$ to the mode $\mathbf{k}$ via the mode $\mathbf{q}$, with $\mathbf{k} = [0, 1, 2]$, $\mathbf{p} = [0, 1, 0]$, and $\mathbf{q} = [0, 0, 2]$ (i.e. transfer of energy to mode with $k_z = 2$ from the most energetic mode). Notably, as $S(\mathbf{k}|\mathbf{p}|\mathbf{q})$ is not zero, even prior to the moment of bursting, there is a discernible energy transfer among these modes. This is likely linked to the waviness in the columns. However, as the system approaches the bursting event, there is a sudden peak in the energy transfer, indicative of a significant perturbation and restructuring of the flow patterns.

Now, from the $S(\mathbf{k}|\mathbf{p}|\mathbf{q})$ of the other triad characterized by $\mathbf{k} = [1, -1, 3]$, $\mathbf{p} = [-1, -1, 2]$, and $\mathbf{q} = [2, 0, 1]$, we observe a distinctive energy transfer pattern. Initially, there is a negligible transfer of energy from the modes with $k_z = 2$ but this becomes noticeable around $t = 94$. Subsequently, during the bursting event, this energy cascade extends to even higher wavenumbers. After the bursting event, the energy transfer pattern returns to that seen before bursting.

In summary, the energy transfer analysis for both triads supports our hypothesis of sequential energy exchange within $k_z$ modes, which intensifies just before bursting, potentially triggering the waviness and the subsequent breakdown of the coherent structures.

### C. Mechanism of bursting

To investigate the cause of bursting and the associated energy transfer, let us first see the $\omega_z$ distribution at the midplane ($z = \pi$) in Fig. 9. In Fig. 10 we show the time-series of aspect ratio, defined as the ratio of the major axis to the minor axis of the elliptical cross-section of the cyclone. Initially, when coherent structures are prominent (Fig. 9(a)), the cyclonic vorticity contours are nearly circular, indicating an aspect ratio close to 1.

Until $t = 92$ (Fig. 9(b)), the interaction between cyclones and anticyclones of unequal strengths (due to the cyclone-anticyclone asymmetry) becomes evident. This interaction is marked by the Crow instability, a well-known phenomena observed for counter rotating vortex pairs [37]. This induces a sinuous deformation of the cyclonic vortex, as seen in Fig. 3(e), similar to the observations by Bristol et al. [38] in their study of unequal-strength vortex instabilities. The Crow instability leads the weaker anticyclonic vortex closer to the stronger cyclonic vortex, causing the anticyclone to wrap around the cyclone and creating a region of intense shear between them. This can be observed in the movie attached as a Supplementary material. The presence of two anticyclones flanking the cyclonic vortex induces a directional stretching effect, resulting in an elliptical cross-section for the cyclone. Up to the time instant $t = 92$, the aspect ratio gradually rises after which we observe a sudden drop perhaps due to an abrupt start of the bursting phenomenon. Thereafter, the aspect ratio rises rapidly before, at $t = 96$ (Fig. 9(c)), the single vortex core fragments into multiple cyclonic regions. The outer cyclonic region becomes even more elliptical, but less coherent with anticyclones encroaching the periphery. The interaction between the cyclones and anticyclones during the bursting phase is very complex yet much intriguing. Near the end of the bursting stage, the aspect ratio reaches its peak value of 1.8. Thereafter, it decreases as the columnar structures start to form again. A similar behaviour is observed for other bursting cycles.

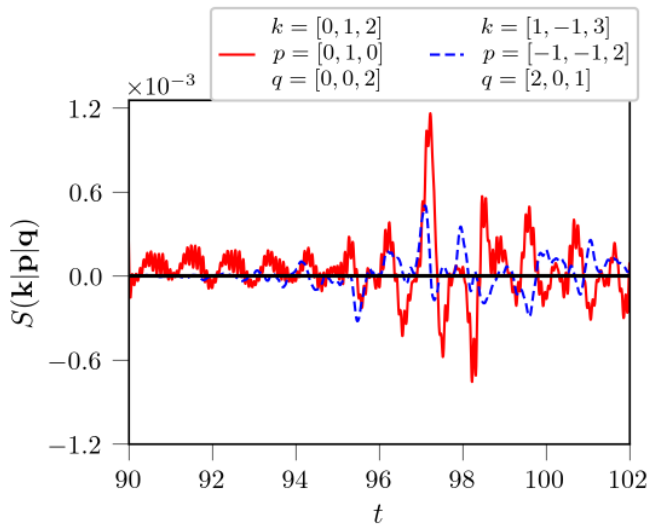

FIG. 8. Time evolution of mode-to-mode energy transfer ($S(\mathbf{k}|\mathbf{p}|\mathbf{q})$) for case 2. Energy is being transferred from mode $\mathbf{p}$ to mode $\mathbf{k}$ via mode $\mathbf{q}$. Dashed line represents the data being multiplied by a factor of 500 for clarity.

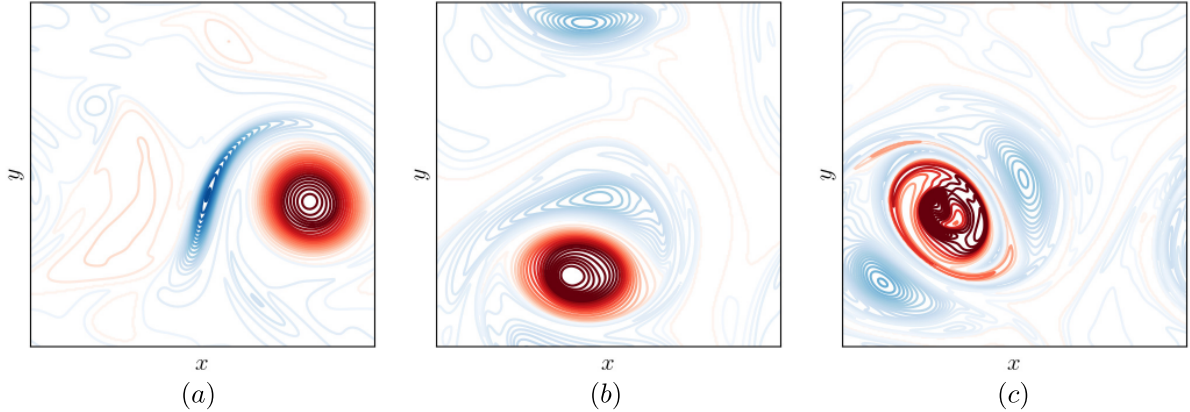

FIG. 9. Cyclonic (red) and anticyclonic (blue) vorticity contours for case 2 in the mid-plane ($z = \pi$) at times (a) $t = 70$ (b) $t = 92$, and (c) $t = 96$. The magnitude of cyclonic $\omega_z$ increases with time, see 3(e, f)).

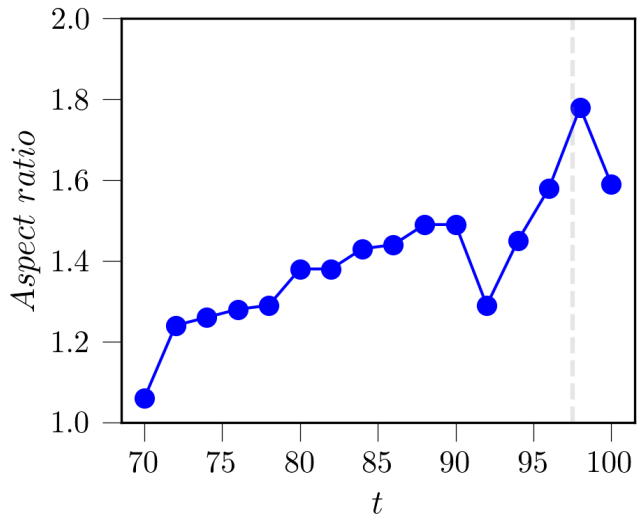

FIG. 10. Time evolution of aspect ratio of cyclonic vortex for case 2 computed at the mid-plane ($z = \pi$) along the direction of rotation. During the bursting the structure breaks down and the aspect ratio represents only the mid horizontal plane. The dashed line indicates the bursting time.

This observation suggests that the elliptical instability, that arises due to a triadic resonance between the strain field from the neighboring vortex and internal wave motions of the affected vortex (Fukumoto [51], Moore and Saffman [52]), is likely to be the mechanism for triggering the bursting. During the bursting cycle, the transfer of energy to modes with $k_z = 2$ appears to reinforce the elliptical instability, leading to a more vigorous vibration of the vortex core. Kerswell [53] predicted that the unstable mode could be destabilized by a secondary instability, potentially leading to turbulent breakdown. A detailed theoretical and numerical analysis of this instability and its growth rate is provided in Appendix A. This amplified elliptical instability, in turn, intensifies the Crow instability, where the cyclone and anticyclone come into contact, increasing the axial flow within the core finally resulting in bursting. In a tug of war, the Coriolis force, acting as a restoring phenomenon, attempts to maintain the cyclonic and anticyclonic structures against this distortion. Thereafter, the vortex forms again due to the background rotation, where the Coriolis force acts as a stabilizing force.

## IV. CONCLUSIONS

In this study, we report for the first time (to the best of our knowledge) bursting of cyclonic columnar structures in simulations of forced rotating turbulence, at low forcing wavenumber. (Our forcing scheme is isotropic, non-helical and injects energy at a constant rate.) This is a complex phenomenon present in simulations at high Reynolds number and small Rossby number. The bursting is sudden and intermittent, and is followed by columnar structure re-formation due to the effect of background rotation. Based on our preliminary investigation into the mechanism of bursting, we suspect that it occurs due to elliptical instability.

Bursting of columns leads to more energy in the higher $k_\parallel$ and $k_\perp$ modes, signifying an energy transfer from $E_\perp$ to $E_\parallel$, opposite to that seen in rotating turbulence [28]. Our analysis reveals the crucial role of $k_z = 2$ modes in the bursting of 3D cyclonic columns, also pointed out by Alexakis [29] for rotating Taylor-Green flow. Higher $k_z$ modes show increased energy during bursting, albeit smaller compared to lower $k_z$ modes. Select modes, such as $[0, 1, 2]$ and $[-1, -1, 2]$, exhibit an increase in energy before bursting, with energy in $k_z = 2$ modes peaking around $t = 97$. This suggests an initial energy transfer to modes with $k_z = 2$, inducing structural waviness, followed by further channeling to modes with higher $k_z$, ending up in the bursting of columns. We suspect that this transfer of energy to $k_z = 2$ modes triggers the Crow instability, leading to a sinusoidal deformation of the coherent column. Continuous transfer of energy to $k_z = 2$ further amplifies this instability driving the weaker anticyclonic vortex to wrap around the stronger cyclonic vortex and creating a region of intense shear between them. This amplified external strain field deforms the core of the cyclone into an ellipse leading to elliptical instability. The coupling of Crow instability with elliptical instability causes a larger axial flow within the vortex core ultimately resulting in bursting. Despite the same forcing mechanism and other parameters, no bursting is observed at higher forcing wavenumbers. A brief comparison can be drawn with rotating convection, where similar columnar vortices are observed (e.g., Guervilly *et al.* [54],Favier *et al.* [55]).However, in those cases, the vortices are long-lived and stable, sustained by buoyancy-driven forcing that is vertically directed and primarily active at smaller scales, resulting in a sustained inverse cascade. This difference in both the directionality and scale of forcing appears to play a critical role in destabilizing the columnar structures, enabling interactions that drive the system toward intermittent bursting. These observations suggests that the nature of forcing plays a key role in whether columnar vortices remain stable or undergo bursting, even in other mechanically forced flows. It is likely that the same bursting mechanism was applicable for the rotating Taylor-Green vortex case by Alexakis [29]. The effect of different forcing mechanisms on rotating turbulence is presently being investigated.

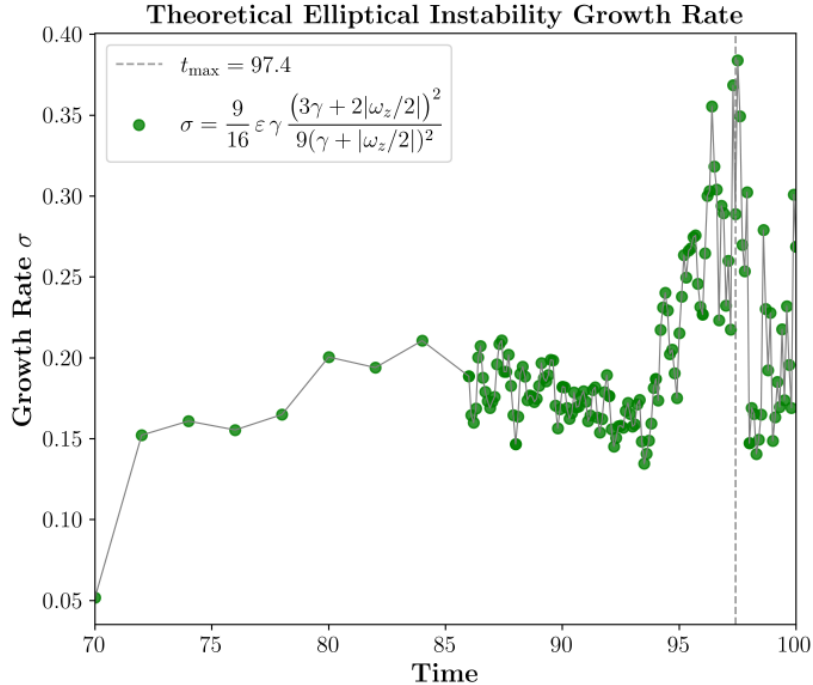

FIG. 11. Time evolution of the estimated elliptical instability growth rate.

**ACKNOWLEDGMENTS**

Authors would like to thank Dr. Anando Gopal Chatterjee and Soumyadeep Chatterjee of IIT Kanpur, for valuable discussions and assistance with the installation and understanding of the pseudospectral code TARANG. AR wishes to thank IRCC, IITB and SERB for the research grants.

**Appendix A: Theoretical and Numerical Estimates of Elliptical Instability Growth**

To investigate the role of elliptical instability in the observed bursting dynamics, we estimate its theoretical growth rate using classical inviscid models. The expression, commonly employed in prior studies (e.g., Le Reun *et al.* [35], de Vries *et al.* [56]), is given by:

$$\sigma = \frac{9}{16}\varepsilon\gamma\frac{(3\gamma + 2n)^2}{9(\gamma + n)^2}, \tag{A1}$$

where $\gamma = \Omega - n$, and $n$ represents the local rotation rate of the vortex core.

Since the vortex cores form coherent vertically aligned columns under the influence of background rotation, the vertical vorticity component $\omega_z$ is much larger than the horizontal components $\omega_x$ and $\omega_y$. We therefore estimate the local rotation rate from $\omega_z$ alone. The vorticity is defined as $\omega_z = \partial u_y/\partial x - \partial u_x/\partial y$, while the antisymmetric part of the velocity gradient tensor (rotation-rate tensor) gives $\Omega_{xy} = \frac{1}{2}(\partial u_y/\partial x - \partial u_x/\partial y)$, leading directly to $\omega_z = 2\Omega_{xy}$. Since $\Omega_{xy}$ represents the local rotation rate about the vertical axis, we approximate $n \approx |\omega_z|/2$, where $|\omega_z|$ is the average vertical vorticity computed over the vortex core.

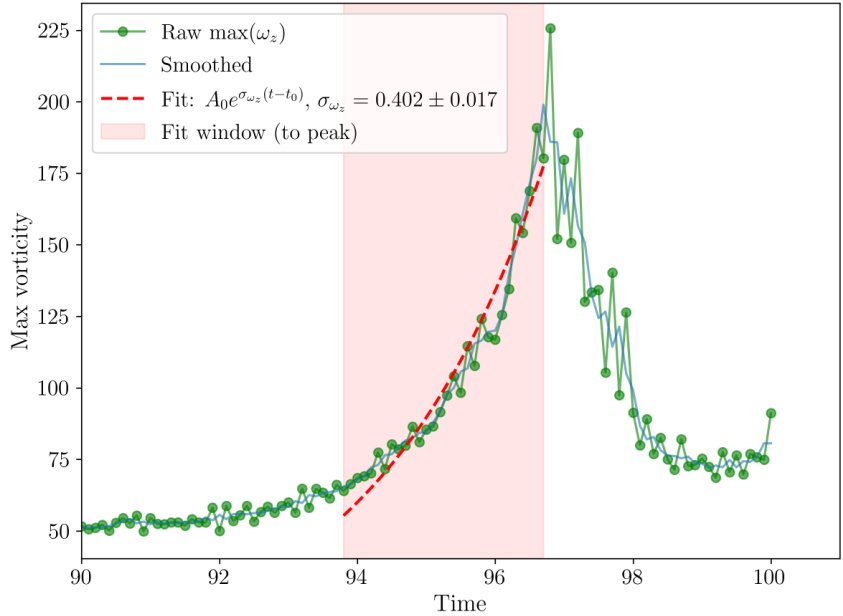

FIG. 12. Exponential fit of max vorticity.

The local ellipticity $\varepsilon$ is calculated from the vortex core geometry using the relation $\varepsilon = \frac{a-b}{a+b}$, where $a$ and $b$ are the major and minor axes, respectively, of the fitted elliptical cross-section of the vortex. These values are extracted from the horizontal vorticity field. The resulting time evolution of the growth rate is shown in Fig. 11. A clear rise in the growth rate is observed around $t \approx 94$–$97$, preceding the burst. It increases sharply from $\sigma \approx 0.20$ at $t \approx 94$ to $\sigma \approx 0.35$–$0.40$ by $t \approx 97$. This trend supports the role of elliptical instability in triggering the observed dynamics.

In addition to the theoretical estimate, we plot the peak value of $\omega_z(t)$ in Fig. 12. An exponential fit is performed over the pre-burst interval $t \approx 94$–$97$, assuming the form $\omega_z(t) \sim Ae^{\sigma_{\omega_z}(t-t_0)}$. This fit yields a growth rate $\sigma_{\omega_z} \approx 0.402 \pm 0.017$.

Since the elliptical instability leads to the amplification of transverse perturbations that ultimately increase the core vorticity, the agreement between $\sigma_{\omega_z}$ and the theoretical prediction $\sigma$ from Eq. A1 provides independent confirmation that classical elliptical instability drives the observed burst.

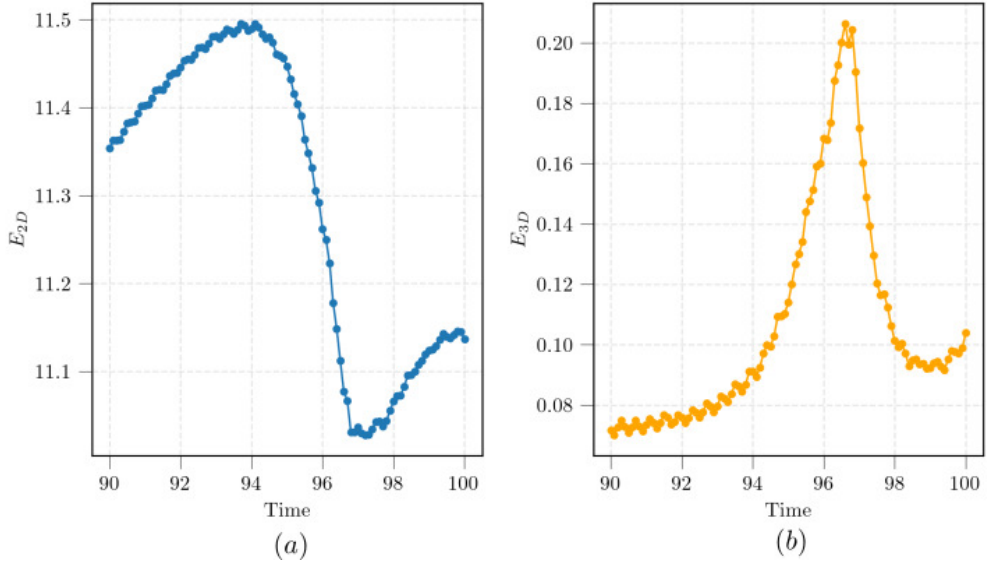

FIG. 13. Temporal evolution of (a) the 2D energy component $E_{2D}$ ($k_z = 0$) and (b) the 3D energy component $E_{3D}(k_z \neq 0)$.

The total kinetic energy is given by $E = \sum_{k_x, k_y, k_z} E(\mathbf{k})$. The energy is separated into 2D and 3D components based on the vertical wavenumber $k_z$. The 2D energy, $E_{2D}$, is computed as the sum of modal kinetic energy over all horizontal wavenumbers at $k_z = 0$ ($E_{2D} = \sum_{k_x, k_y} E(k_x, k_y, k_z = 0)$), while the 3D energy, $E_{3D}$, is obtained by summing over all modes with $k_z \neq 0$ ($E_{3D} = \sum_{k_x, k_y, k_z \neq 0} E(k_x, k_y, k_z)$).

In Fig. 13, we show the temporal evolution of $E_{2D}$ and $E_{3D}$. During the burst phase, $E_{3D}$ increases, while $E_{2D}$ decreases, consistent with the growth of elliptical instability and the breakdown of vertically aligned columnar structures. This trend is also evident in the analysis of the ring spectrum presented in Section III B (Fig. 6), where the enhanced energy appears in the larger $k_{\parallel}$ during the bursting interval. The evolution of modal energy $E(k_z, t) = \sum_{k_x, k_y} E(k_x, k_y, k_z, t)$ for the first few vertical modes ($k_z = 1, 2, 3$) is shown in Fig. 14. Each curve exhibits a clear linear trend on a semi-logarithmic scale, indicating exponential growth. Fitting the form $E_{k_z}(t) = B\, e^{\sigma_{k_z}(t-t_0)}$, yields growth rates $\sigma_{k_z=1} = 0.355 \pm 0.005$, $\sigma_{k_z=2} = 0.425 \pm 0.011$, and $\sigma_{k_z=3} = 0.465 \pm 0.015$.

These mode-specific rates cluster around the vorticity-based estimate $\sigma_{\omega_z} = 0.402 \pm 0.017$

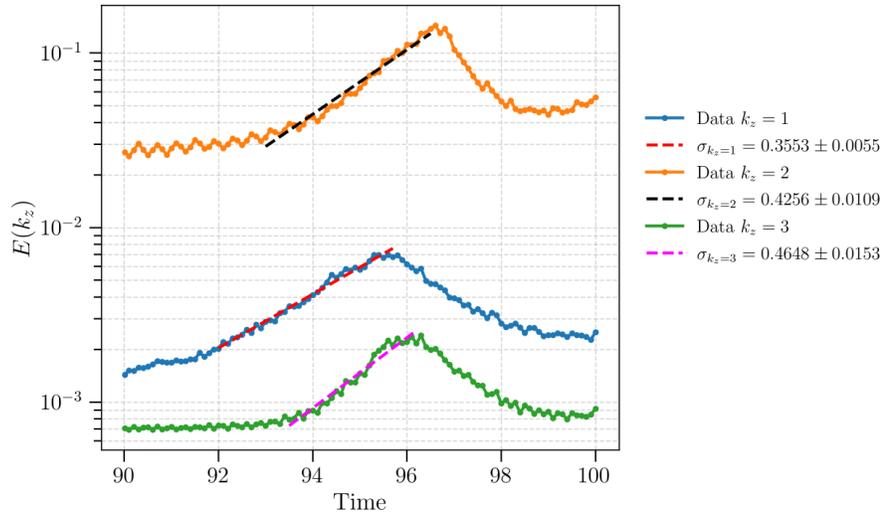

FIG. 14. Temporal evolution of the vertical mode energies $E(k_z)$ for $k_z = 1, 2, 3$. Exponential fits are applied to each curve to extract the corresponding growth rates $\sigma_{k_z}$ during the pre-burst phase.

and lie close to the theoretical range of 0.35–0.40, demonstrating that the same elliptical instability is responsible for both the amplification of vortex cores and the transfer of energy to 3D modes. We also observe that $\sigma_{k_z}$ increases with $k_z$, and that exponential growth for each mode begins progressively later for higher $k_z$, suggesting a delayed activation of smaller vertical scales. This energy growth is correlated with transfers to specific modes such as $[0, 1, 2]$ and $[1, -1, 3]$ from $k_z = 0$ modes, as shown in the mode-to-mode transfer analysis in Fig.8.

---